# Analytical review of medical mobile diagnostic systems

D. Kordiyak[1], N. Shakhovska[2]

[1]*Lviv Polytechnic National University; e-mail: d.kordiyak@gmail.com*
[2] *University of Economy, Bydgoszcz; e-mail: natalya233@gmail.com*

*Abstract.* This article analyzes the mobile medical diagnostic systems and compare them with the proposed HealthTracker system based on smart watch Apple Watch. Before the development of the system HealthTracker, there was conducted a review and analysis of existing similar systems to identify common and distinctive features of the future system. This analysis will improve HealthTracker system, based on the strengths and weaknesses of existing systems and help identify and justify the key benefits and unique system HealthTracker. The main goal is to provide a system HealthTracker convenient way to interact with the patient the doctor based on the vital signs of the patient. Apple Watch is an excellent watch presented in 2014 that has the capacity to collect and compile data on the health of the user and can be used for medical purposes. The main hardware components of the watch for collecting and analyzing health data is a technology Taptic Engine, infrared sensors and pulse. The main software components of the watch, that will be used in the design of the system is the 3 applications, each of which measures a user's vital signs. Integration with smartphone user makes data on the health of a quick and reliable. On the market today there are analogues of the system, but most of the systems are relatively new and require many improvements, some are under development prototypes. In addition, all the above systems require binding to certain equipment that is not always convenient in everyday use. To eliminate all the inconvenience in using existing systems need to create a system that is integrated into smart watches that provide ease of use, and the mechanism storing and analyzing medical data to cloud storage. An important aspect of the study is to analyze the general situation in the market of mobile medical diagnostic systems. Thanks to research the key advantages and disadvantages of the proposed mobile medical analysis system and shows its versatility compared with existing systems on the market

## INTRODUCTION

The modern world is moving towards globalization in all of it's key processes and phenomens, ranging from socio - economic to cultural processes. For information technology the globalization process is the fastest. More and more functions that were traditionally performed by personal computers, are transferred to mobile devices because people want to read and work with their digital data at any moment anywhere. Recently, the cloud technologies became widely spread because of carefully universal access and processing of data. Cloud technology in many times simplified the process of common access to several workstations data. The traditional expert systems gradually fade into the past because of the many exceptions to the rule and large databases. Instead of them, mobile expert systems are becoming widely and widely spread. However, there is a certain dependence on platforms, integrating data from various devices and more. Depending on the goal of developing a particular system one should choose the most suitable platform and work with the appropriate types of devices.

The main purpose and objective of the program system is to collect patient medical data directly without the use of special medical equipment and devices. Today becoming more common so-called "smart" watches. The main difference between these clocks from conventional counterparts is the ability to connect to the Internet and synchronization with a smartphone. In addition, some models have hardware and software that allows you to collect patient medical data (user). These models of watches have special sensors that provide a variety of human medical screenings, such as heart rate, pressure, etc. With these impressions doctor can monitor the patient's condition in real time and on time and quickly stanovlyuvaty patient diagnosis and provide recommendations.

As the clock, which will be used in developing I've chosen Apple Watch. This is one of the most functional "smart" watches today and its integration with the iOS are pretty well thought out.

Before the development of the system, I've conducted a review and analysis of existing similar systems to identify common and distinctive features of a future system. This analysis will improve the system HealthTracker, based on the advantages and disadvantages of existing systems, and will detect and justify the key benefits and uniqueness of the system HealthTracker. It is necessary to analyze the existing system, they identify unsolved problems in order to solve them in your system.

Many fitness - trackers on the market is relatively effective in mild calculations such as counting steps, but in more severe problems, such as heart rate measurements even approximate accuracy mostly lost. In addition, the question of the value of clinical data provided fitness - trackers because the number of steps and heart rate data on the user not processed. It is this question solves the following system.

## THE ANALYSIS OF RECENT RESEARCHES AND PUBLICATIONS

The newest similar system that was introduced in mid-2015 exhibition International Modern Hospital Show in Tokyo, is a system Silmee. This intelligent sensor measuring parameters of human life from Toshiba. This sensor is the ability to remotely monitor key indicators of the body, which is important for patients, pilots, rescuers and ordinary people who are constantly worried about the state of their loved ones. This sensor simultaneously collect data on body: does electrocardiogram measures the heart rate, body temperature, record the motion of the media, etc. These data can be transferred over a wireless connection, such as Bluetooth, to your smartphone or tablet, and then - at any point on the planet. The new sensor is equipped with 32 - bit chip, wireless data transmission system. The collected sensor will have dimensions of 2.5 to 6 cm and weighs only 10 grams. Such small size and weight allow you to use it in a variety of applications, from monitoring the patient to postoperative surveillance of soldiers during training.

The basis of the system is Silmee two prototypes bracelets - Silmee W20 and Silmee W21. Modification additional W21 is equipped with GPS, which is especially useful for the elderly. This option allows relatives ever see the place where they are based on their smartphones. In the body of the unit taken out button SOS, because of this, when you click on the device is passed through the smartphone alarm in the medical institution or ambulance services.

In addition, Toshiba has shown sensor Silmee Bar type Lite. This device is secured to the chest in ADR for helium pads and allows to measure respiratory rate.

Detailed specifications of devices:

Toshiba Silmee W20 - wrist sensor having classifications IPX5 and IPX7, dimensions 20.5 x 65 x 12 mm and weight of 29.5 grams. The device is equipped with a lithium - ion battery, which in real terms is enough for 2 weeks. The cost of the device - $ 193.

Toshiba Silmee W21 different physical dimensions 25.5 x 65 x 12.5 mm, weighing 38 grams and the availability of GPS module. The cost of the device - $ 225.

Both devices are compatible with iOS version 7 or higher and Android 4.4 and above.

The advantages of this system is the high functionality, the large number of sensors that provide a lot of performance and price of human life in general. The presence of two versions allows you to select a suitable option for a particular user. In addition, the physical parameters of the device, such as dimensions, is relatively good compared to the competition.

The main disadvantage of the system is awkward to use software on a smartphone due to the relative newness of the system and not very thoughtful UI and UX. In order to use the system need to buy one of braceletes which is not very convenient for people using the watch. [1]

## OBJECTIVES

The main objective of the research is to highlight most innovative and popular medical systems, investigate them and conduct main features of HealthTracker system.

## THE MAIN RESULTS OF THE RESEARCH

1.1. iHealth Products

Another system that will be considered is the iHealth system from the same named company, developed in Silicon Valley in the US in 2009. In 2013 the company opened an office in Paris - iHealthLabs Europe. iHealth develops and distributes innovative, clinically tested products in the health sector.

The company's products pass rigorous testing before be issued on the conveyor. Blood pressure monitors are tested protocol ESH (European Society of Hypertension), blood glucose meters tested the protocol ISO15193: 2013.

Data on vital functions of the patient is stored on the cloud. The company has developed a series of special cloud data stores, which are regulated by international standards of data security. When a user synchronizes with each product with your smartphone or tablet, information is sent to one of the clouds. With this cloud storage, the user has access to their medical data from anywhere - any computer, smartphone or tablet at all - anywhere in the world. Medical patient data is completely anonymous and confidential, but at the request of the patient, these data can be sent to a doctor for the rapid analysis of health.

The company provides a choice of nine different products.

*Connected Wrist Blood Pressure Monitor* – the monitor that measures blood pressure, pulse and is able to detect arrhythmia. The device is placed on the wrist of the patient. Constant monitoring of patient's pressure is an important issue in medicine. This device allows the doctor to monitor blood pressure and heart rate to better conduct treatment for the patient and offset the cardiovascular risks. The device comes with iHealth MyVitals application that provides a simple user interface with a variety of graphs and tables. This application can continuously monitor blood pressure according to the standards of international public health data and to share these with your doctor. Due to the powerful battery device allows almost 80 times to measure and save data on patient pressure. [2]

*Connected Arm Blood Pressure Monitor* – similar to the previous device, difference is that ut's not placed on the wrist, but on the hand. In addition, the device is 20% more expensive than the previous. [2]

*Connected Scale* – smart weight, allowing the patient's weight regularly monitored to prevent excess weight gain or loss. In addition, these weights allow to measure the effects of a balanced diet and regular physical activity. As usual weight, the device is instantly and accurately reflects the weight of the user, but by connecting to the mobile application iHealth MyVitals App, the product allows the user to accumulate statistics scales, save them to cloud storage and send data to the physician. It stores up to 200 measurements in memory thus saved great history of previous measurements of gravity that allows the patient to improve the quality of statistics. [3]

*iHealth Wireless Body Analisys Scale* – similar to the previous measures, but also calculate the index of body fat, bone mass, muscle mass and water weight in the body. Thanks to more parameters, the user is much easier to pick the correct and balanced diet and learn about the state of the organism as a whole. Compared with the previous scales, the product is more expensive by 30%. [3]

*Connected Body Analisys Scale* – weight that combine the two previous products, but also with a special algorithm to assess the optimal number of calories that the user should be taken in the near future for better health. This product is the latest product of the company and is expensive from the base weights of 40%. [4]

*Activity & Sleep tracker* – this device can be worn all day long, it measures the number of steps, distance traveled and calories burned during the day. Through synchronization with the corresponding application, one has the opportunity to check their results, view statistics and set a new goal in the number of passed steps (day norm is 10,000 steps). Bonus functionality of the device is silent alarm. [4]

*Wireless Pulse Oximeter* – device to monitor oxygen levels in the blood and pulse. This simple and reliable device allows you to monitor the level of oxygen in the blood, pulse and perfusion index. These measurements will be useful for people with chronic diseases such as asthma or obstructive disease pulmonarnoyu to fast track their health. In addition, the device will be useful for smokers that quit smoking because they constantly monitor the improvement of blood and it gives them motivation. Athletes will see the body's response to certain types of exercise. Due to the synchronization of the application, user can observe the progress of chronic diseases or the result of exercising. The device is extremely easy to use, just put your finger on it and in seconds you'll have the results. [4]

*Connected Mini Glucometer* – compact handheld device to measure blood sugar levels. The device is useful for people with diabetes to monitor blood glucose levels. On the output device gives blood glucose levels, writes them to the internal memory, and refers to cloud storage to continually inform your doctor. Also, there is the possibility of establishing notifications under certain limits blood sugar. The device connects to the smartphone via AUX - connector, it provides high portability. [4]

*Wireless Smart Gluco – Monitoring System* – device that has similar functionality to the preceding except much worse and fully informing the patient of data on blood sugar levels. Availability of algorithm that predicts changes in blood sugar levels allows conduction of better treatment of patient. [5]

The advantages of iHealth products is it's high accuracy and reliability, long experience on the market. A wide range of products and components makes the company an influential player on the market. A small price products ensures their availability to the end user. [5]

Disadvantages of iHealth products are too high differentiation and lack of versatility, use as many products, most of which is in constant use light is uncomfortable. In addition, technical devices are built so that in case of malfunctions the user is able to disassemble and repair their own. Also, products are not hit and water proof making them uncomfortable to use in the real world.

1.2. Another existing products

Another similar system is AiQ BioMan. The company was able to put a mobile medical system to an ordinary T-shirt. This is the ability to measure levels of heart rate, breathing and skin temperature. In addition, the more expensive versions of the product are sensors electrophysiological signals. The advantages of the product are innovative solutions , the drawback - low accuracy of measurements and their lack of analysis or sync with the repository. [5]

*Metria Wearable Sensor* is a compact device that is attached to the body and collects data on the number of breaths per minute, the number of hours of sleep manual. With WiFi, device sends the data to the smartphone user. The advantage of the device is the low price and the possibility of use in many activities, the lack is a small number of indicators and the lack of medical analysis results. [5]

*BodyTel* a set of devices that include measuring blood glucose levels and pressure. Each device has a Bluetooth - module, through which sends the data to the home station user. Upon receipt of the required data, the station sends data to a database that is available to the physician. In addition to viewing data, doctors can send notifications to patients based on restrictions of certain indicators that determined by the physician. Home station also provides feedback to a physician when one or more indicators of the health of the patient out of the limits of normal. The advantages of the system are established relationship with a doctor and thought-out system alerts, disadvantage is the attachment system to the home station, making it impossible to mobility system. In addition, sending data to the home station is only possible via Bluetooth. [7]

*Imec* – is a mobile electroencephalograph to monitor the state of the brain of the patient. All the data is automatically sent to the smartphone user. This system is currently under development and is a condition of existence prototype. The system is still incomplete and did not pass the necessary tests to entry. [6]

*Moticon* – a tech supinator, which works with the parameters of the movement of the carrier. The device measures the number of steps carrier weight and other parameters. The advantage is the ability to use in any type of shoe. The disadvantage of the device is a connection with a computer only via USB - Beacon which is not very convenient to use and allows you to get the required data in real time. [8]

*Preventice BodyGuardian RMS* – system for continuous monitoring of the health status of patients with heart arrhythmia. It allows the doctor to view patient's required performance thanks to the small sensor that attaches to the chest of the patient. Then smartphone data is sent to the physician. The advantages is high accuracy results, the drawback is the complexity of use [9].

2. Apple Watch overview

Apple Watch is a smart watch, designed by Apple Inc. It includes a fitness track and the possibility of reviewing the state of health as well as integration with iOS and other products and services Apple. The device is available in four versions: Apple Watch Sport, Apple Watch, Apple Watch Hermès, and Apple Watch Edition. Watches vary with different materials body and screen size. Apple Watch is based on wireless connected iPhone to perform many of its functions by default, such as the integration of call and offers a set of text messages. It is compatible with iPhone 5 or later models running iOS 8.2 or later, through the use of Bluetooth or Wi-Fi. The device was announced by Tim Cook on Sept. 9, 2014. Apple Watch quickly became a bestseller from the release of more than 4200000 units in the second quarter of 2015, according to research firm Canalys. [10]

The aim of a device was to free people from their phones. Apple Watch works when connected via Bluetooth or Wi-Fi to your phone and then using it to access all applications that are downloaded to the mobile device. Development Apple Watch took place in secret and was relatively unknown until then, until the article was published Wired, which can be found online. [11]

The hardware part. In Apple Watch uses the system on a chip - S1. It uses linear drive called "Taptic Engine" to ensure connectivity tactile feedback upon receipt of notification or communication and is used for purposes other relevant applications. Clock features a built-in heart rate sensor, which uses both infrared and visible spectrum of light and photodiodes. All versions of Apple Watch clock with 8 gigabytes of memory, the operating system allows the user to store up to 2 gigabytes of music and 75 megabytes of photos. The main factor why it was chosen Apple Watch in the development of the system is that it can create a pair with the iPhone and sensors transmit data to the program Apple Health, through which you can get the necessary information about the patient's condition. Battery of Apple Watch maintains more than 18 hours of mixed use. Clock charged via inductive charging cable, which is similar to the MagSafe from the MacBook. If the battery is less than 10%, the clock will switch to power saving mode that allows you to see just an hour. In addition, the product has waterproof elements, it potentially allows you to monitor the physical condition of swimmers. Special tests have shown that Apple Watch can operate in conditions of immersion under water in different conditions, though the sensor then becomes less accurate. [12]

The software part. Apple Watch runs on OS - WatchOS, whose interface is based on the home screen with circular icons set programs. Moving the home screen takes place by means of sensors or corona side of the clock. The watch is able to receive messages and calls using paired iPhone. "Views" allow users to switch between pages containing widgets. WatchOS supports Handoff, to send content from Apple Watch for iOS or OS X and used Siri voice commands. Apple Watch also supports Apple Pay, and can be used with older models of iPhone, do not support NFC. Applications of Apple Watch are designed to interact with iOS - analogues, such as e-mail, phone, calendar, messages, maps, music, photos, reminders, etc. With the development of HealthTracker system, the application we are interested in is Fitness app that keeps track of physical activity of the user and sends the data back to the iPhone Health App. WatchKit - application runs in the background of the iPhone as an extension applications and send data about the physical condition of the media to the phone. Thus, WatchOS - applications must be paired with the iOS - application and automatically sync with iPhone. [13]

In the future Apple Watch will use an operating system WatchOS 2, was presented at WWDC 2015. It added support for native applications, added night mode, new dials with customizable time - lapse video as the background, and many other changes. [14]

Also there are new opportunities for programmers in WatchOS 2. The main innovation is a native SDK, which improves parameters benchmark programs, because they are written by native WatchOS 2. In addition, developers granted direct access to the accelerometer, microphone, crowns, and moreover, to medical devices and sensors TapTic Engine, that will be used in the development of the system.. [15,16]

Apple Watch was selected for appropriate device for health tracking for next reasons:

- The display device is extremely sensitive. Apple has been able combine small screen sizes with very accurate sensitivity of the screen. This makes it easy for navigate throughout the screen and choose appropriate programs and controls. In addition, the display reacts differently to the pressure force. Also Retina - display provides high clarity of screen elements.

- The main radial menu. All the programs are of the screen and scroll works in all directions.
- High stock battery. Apple claims that the battery keeps watch about 18 hours of active use, various tests have shown that by the end of the day is about 25% of the battery, which is a fairly good indicator in comparison with analogues.
- Built-in integration with Siri. This technology allows voice commands to perform a great variety of tasks without requiring user attention, including Dictation typing, content tagging, ratio reminders and searches on the Internet.
- Ability to answer the phone. Currently it's the only clock that allows direct incoming calls when paired with a smartphone.

Despite the considerable number of advantages clock has its drawbacks.

The main disadvantage is that without connection with the iPhone, most of features of the watch are nullified, including the ability to track health, a key necessity in the development system. [17]

Another drawback of the device is its price. The price varies depending on the type of clock. In developing the system will be used and Apple Watch Apple Watch Sport with displays 38 and 42 inches respectively, watches materials are aluminum and stainless steel. Apple Watch Edition is an exclusive product with a price of $ 10,000, which does not mayuye no advantage in terms of functionality and differ only cost materials (18 - carat gold). Apple Watch Sport price starts from 349 $, Apple Watch from $ 599. But in the increase in the dollar price of such a device are considered a disadvantage.

The third potential drawback of the device is that due to changes in color, such as tattoos, working sensor device is inaccurate, making it impossible to establish the exact state of health of the patient. The reason is that monitoring clock pulse user svityst green flash on the skin and recording the number of green light that is absorbed by the blood. However, after the change of color, this functionality can not fully correct opratsovuvavty rates of absorption of green light, causing possible inaccuracies in the results. [18]

One of the main applications, which is useful in exploring the physical condition of the user and is used to design the system is Activity App. This application provides a simple visual representation of daily physical activity guide with three rings, showing all relevant information. The so-called "Stand ring" shows how often people got up from their seats. "Move ring" indicates the number of calories burned by the user. "Exercise ring" shows how many minutes of exercise done. The main purpose of this program is the ability to install your daily limit individually and closing these rings. [19]

Stand. Sitting too long has some health risks. Apple Watch Sensors determine when the user stands up and record it. If you sit for more than an hour, the clock is reminiscent of the media. "Stand ring" closes when the user gets up and moves at least 1 minute every hour of the day. This may seem trivial, but frequent breaks from sitting a positive effect on health in general.

located radially from the corners

Move. Every week, Apple Watch offers a new goal in terms of traffic every day to burn some calories, given recent indicators user. "Move ring" completes when the user closes the established amount of burnt calories for the day. The user can change the proposed amount of calories, increasing or decreasing it depending on the state of health of a day. The main purpose of "Move ring" is a small but constant improvement in traffic every week for progress of health of the user.

Do the exercise. Apple Watch stores data on physical activity every day user. "Exercise ring" closes when the media do at least 30 minutes of exercise every day. Exercise is not necessary to perform row for half an hour, the user himself distributes their number during the day depending on their own recommendations.In addition to these general functions of the program Activity App, there are many other applications built by different companies to promote their products. With these applications, the clock will be useful for cyclists, people do yoga or gymnastics. [20]

CONCLUSIONS

Today there are a large number of mobile medical systems that measure a variety of indicators of human activity. However, most of the systems are relatively new and require many improvements, some of them are just development prototypes. In addition, all the above systems require binding to certain equipment which are not always convenient in everyday use.

To reduce inconvenience in the use of existing systems we need to create a system that is integrated into the smart watch that will provide ease of use, and the mechanism to store and analyze medical data to cloud storage.

With all bound by these clock capabilities, there can be made a conclusion about the suitability of its use in medicine, probably the best device for recording each change in the physical activity of the patient and establishing an early diagnosis. The system integrates directly with the clock for convenience and requires no additional devices or settings. All the carrier's physical performance will be processed and stored system in the cloud, which will have access to a doctor. This will help to maximize the effectiveness of the system in real conditions.